\documentclass[12pt]{article}
\usepackage{epsfig,graphics}

\newlength{\figsize}
\begin{document}

\begin{titlepage}

\begin{tabbing}
\` Oxford preprint OUTP-02-28P \\
\end{tabbing}
 
\vspace*{0.7in}
 
\begin{center}
{\large\bf The deconfinement transition in SU(N) gauge theories \\ }
\vspace*{0.5in}
{Biagio Lucini, Michael Teper and Urs Wenger\\
\vspace*{.2in}
Theoretical Physics, University of Oxford,\\
1 Keble Road, Oxford, OX1 3NP, U.K.\\
}
\end{center}

\vspace*{1.25in}

\begin{center}
{\bf Abstract}
\end{center}

We investigate the properties of the deconfinement 
transition in SU(4) and SU(6) gauge theories. We 
find that it is a `normal' first order transition
in both cases, from which we conclude that the transition
is first order in the $N\to\infty$ limit. Comparing our 
preliminary estimates of the continuum values of $T_c/\surd\sigma$
with existing values for SU(2) and SU(3) demonstrates a weak dependence 
on $N$ for all values of $N$.

\end{titlepage}

\setcounter{page}{1}
\newpage
\pagestyle{plain}

\section{Introduction}
\label		{intro}

The dependence of QCD and SU($N$) gauge theories on the
number of colours, $N$, is of great interest
\cite{largeN,mtN}.
Through a number of recent lattice calculations 
\cite{mtN,blmt-kstring,blmt-glue,Pisa1,Pisa2,ncmtuw}
we have learned a great deal about the mass spectrum,
various string tensions and the topological properties
of SU($N$) gauge theories at finite and infinite $N$.  
These calculations confirm the usual diagrammatic expectation 
that a smooth large-$N$ physics limit is achieved by keeping 
fixed the 't Hooft coupling $g^2N$. In addition there  
is some good evidence from very accurate calculations of the 
string tension, $\sigma$, in 2+1 dimensional gauge theories 
\cite{d3},
that the leading 
corrections are $O(1/N^2)$ -- again as expected. (Note 
however that there is room for a violation of this
\cite{mtN},
but the evidence is equivocal as yet.) 

An interesting question concerns the order of the deconfining
phase transition at $N = \infty$. It has been argued
\cite{pisarski1}
that the difference between the SU(3) gauge theory and QCD 
with light quarks (see
\cite{karschrev}
for recent reviews) can be reconciled with both being 
`close to' large-$N$ if in fact the  SU($N$) transition 
becomes second order at large-$N$. (See
\cite{pisarski2}
for some more recent reflections on this.) This has prompted 
lattice studies
\cite{winoh,gavai}
of deconfinement in SU(4) which claim that the transition
is first order there. However it is only in SU(6),
or larger groups, that all possible escape routes close
\cite{pisarski1}.
Thus we hope that the calculations in both SU(4) and SU(6) 
gauge theories
\cite{blmtuw}
that we shall briefly report upon in this paper will
serve to settle this question. (See also
\cite{campostrini}
for calculations at very large $N$, performed in a 
twisted Eguchi-Kawai single-site reduced model
\cite{TEK}.)

In this letter we focus upon the order of the transition and 
upon our preliminary determination of how the critical
temperature, $T_c$, varies with $N$.
There are of course many other interesting aspects
of the transition, some of which we are studying.
We will mention some of these in the concluding section,
and will present our detailed results in a longer publication
\cite{blmtuw}.
There we will also discuss and compare several different
methods for locating and analysing the deconfining phase
transition. We will, in addition, apply these methods to the 
case of SU(2), where the transition is known to be 
2nd order 
\cite{Tsu2},
and SU(3), where it is known to be weakly 1st order 
\cite{Tsu3}.
At the same time we will have higher statistics which should 
allow us to considerably improve upon the preliminary continuum
values of $T_c/\sqrt\sigma$ that we present in this paper.
However, we believe that we already have unambiguous evidence 
for the first order nature of the SU(4) and SU(6)
phase transitions and this is what we wish to 
communicate in this letter.

\section{Strategy and technical details}

Our lattices are hyper-cubic with periodic boundary conditions
and with lattice spacing $a$. We use the standard plaquette action 
\begin{equation}
S = \beta \sum_{p}\{1-{1\over N}{\mathrm {ReTr}} U_p\}
\label{B1}
\end{equation}
where the ordered product of the SU($N$) matrices around 
the boundary of the plaquette $p$ is represented by $U_p$.
In the continuum limit this becomes the usual
Yang-Mills action with
\begin{equation}
\beta = {{2N} \over {g^2}}.
\label{B2}
\end{equation}
The simulations are performed with a combination of heat bath 
and over-relaxation updates, as described in 
\cite{blmt-glue,blmt-kstring}.
Our SU(4) and SU(6) updates involve 6 and 15 SU(2) subgroups
respectively.

Consider a lattice whose size is $L^3 L_t$ in lattice 
units. This may be regarded as a system at temperature
\begin{equation}
a(\beta) T = {1\over{L_t}}.
\label{B3}
\end{equation}
We require $L \gg L_t$, so that it makes sense to discuss thermodynamics.
We study the deconfining transition by fixing $L_t$ and varying 
$\beta$ so that $T$ passes through $T_c$. 

For $T<T_c$ we are in the confining phase, while for  $T>T_c$
we will be in one of $N$ equivalent deconfined phases.
To distinguish all these phases we define the Polyakov loop
\begin{equation}
l_p({\vec x}) = {\mathrm {Tr}} \prod_t U_t({\vec x},t)
\label{B4}
\end{equation}
and denote by ${\bar l}_p$ its average over ${\vec x}$ for a given
field configuration. For $T<T_c$ we will have ${\bar l}_p \simeq 0$
while for $T>T_c$ the $N$ deconfined vacua will be characterised
by values ${\bar l}_p \simeq c z_n$ where $z_n$ is one of the $N$'th 
roots of unity and $c > 0$ will depend on $\beta$, $L_t$, $N$ etc.
For $L$ large enough the fluctuations of ${\bar l}_p$ become small
enough that a field configuration can be unambiguously 
categorised into one of the phases by its value of ${\bar l}_p$.
Near $T_c$ there may be tunnelling between these phases.
This process will involve field configurations that contain
a mix of phases and will have intermediate values of ${\bar l}_p$.
Note that the usual characteristic of first order lattice
transitions, a clear double peak structure when one plots
a histogram of the number of fields against the action (energy)
is not so useful here. This is because we are dealing with a
physical rather than a bulk transition. Any difference in
the action will therefore be $O(a^4)$ and this is extremely
small compared to the $O(1/\beta)$ value of the action. It
is only on the very largest lattices that the fluctuations
in the latter, which are also $O(1/\sqrt{L^3})$, become small
enough for the beginnings of a double peak structure to
appear. By contrast, clear multiple peak structures in histograms
of  ${\bar l}_p$ already appear on small lattices. 

In a finite spatial volume $V$ the phase transition is smeared
over a finite range of $T$. One needs to choose a value for 
$T_c(V)$ in this range and then to extrapolate this to $V=\infty$ 
to obtain the value of $T_c$ in the thermodynamic limit. 
Different criteria will differ at
finite $V$ but should extrapolate to the same value of $T_c$ at
$V=\infty$. Good criteria will be ones with finite-$V$ 
corrections that are simple, small and have a known
functional form. We have performed calculations with
a number of criteria which produce consistent results.
In this paper we present values of $\beta_c$
obtained from the value of $\beta$ at which the Polyakov loop 
susceptibility has a peak. To obtain the susceptibility
as a continuous function of $\beta$ we use a standard
reweighting technique
\cite{reweight}
applied to those calculations that lie close enough to $T_c$. 
This will be described in some more detail in the next Section.

Tunnelling between the confined and deconfined vacua is
a first sign that the phase transition is first order. 
However its observation at $T\not= T_c$ and/or on small
volumes may be misleading. Equally a lack of tunnelling
on small volumes may be misleading; for example,
the transition may be weakly first order. It is therefore
crucial to perform a finite volume study to identify
what volumes are large enough to be useful. Given that
near $T_c$ the important length scale is $aL_t = 1/T_c$
one would expect that one needs at least $L \simeq 3L_t$
in order that the periodic spatial volume can accommodate 
two phases and some intervening surface in a realistic
fashion. In a finite volume study a first order transition
would be characterised, at large $V$,  by a latent heat
that stays non-zero, a mass gap that does not vanish
(i.e. a correlation length that does not diverge),
an appropriately behaved susceptibility, and
clear tunnelling that becomes exponentially rare
as $V$ grows.

The computational cost will clearly be minimised by using
the smallest possible values of $L_t$. It is however well
known that with the standard plaquette lattice action 
there is a cross-over in the
strong-to-weak coupling transition region which becomes a
strong first-order phase transition at larger $N$. (See e.g.
\cite{blmt-glue}
for a discussion.) For SU(4) this appears to be smooth
cross-over 
\cite{blmt-glue}
but by SU(6) it is certainly a strongly first order bulk 
phase transition
\cite{Pisa1}.
This presents some problems. In particular any calculations
that are going to be useful for continuum extrapolations
will need to be made on the weak-coupling side of this
transition. For SU(4) deconfinement this means that we
have to use $L_t \geq 5$. For SU(6) the finite-$T$ transition 
for $L_t=5$ is quite close to the strongly first order bulk phase 
transition and so this would not be a good place to address the
delicate question of whether the former is also first order
or not. For that we will use $L_t=6$. 
However the computational expense of SU(6) calculations
at $L_t=6$ is such that we will not be able to perform
the kind of finite size study that distinguishes most unambiguously
between 1st and 2nd order phase transitions. Fortunately
the phase transition turns out to be reasonably strong first 
order, rather than being weak first order as in SU(3). 
This makes it relatively easy to study, even within our
tight constraints.

Our strategy will therefore be as follows. We start with
SU(4) where we perform a detailed finite size study of
the deconfining transition for lattices with $L_t=5$.
We shall establish that it is first order and will
quantify its dependence on the spatial volume. We
then perform a study of the SU(6) phase transition on a 
$16^3 6$ lattice, using our earlier finite-size study to 
extrapolate it to the thermodynamic limit. We also have begun
calculations in SU(4) for $L_t=6$ and in SU(6) for $L_t=5$;
these enable us to make some preliminary estimates of
$T_c/\surd\sigma$ in the continuum limit. At the same
time the $L_t=6$  SU(4) calculations provide for
a direct comparison with the SU(6) calculations at $L_t=6$.

\section{$T_c$ in SU(4)}

\begin	{figure}[p]
\begin	{center}
\leavevmode
\epsfig{figure=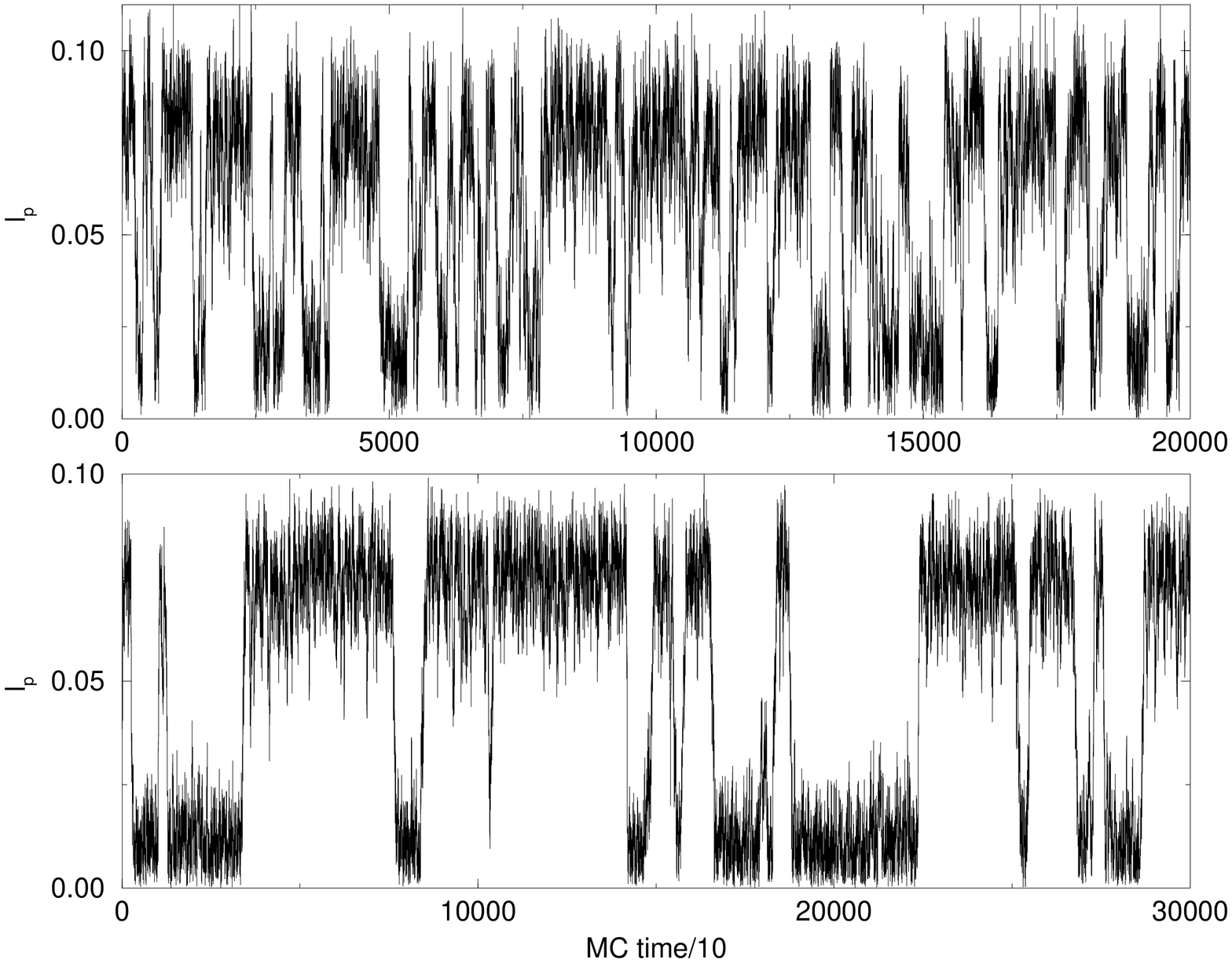, angle=270, width=15cm} 
\end	{center}
\vskip 0.15in
\caption{The modulus of the average value of the Polyakov loop 
for a sequence of $14^3 5$ (top) and $20^3 5$ (bottom) field 
configurations at $\beta=10.635$ in SU(4).}
\label{fig_history4}
\end 	{figure}

Our $L_t=5$ study has been performed on $L=12,14,16,18,20$
lattices, with between 200000 to 300000 sweeps at each value
of $\beta$ and $L$. For $\beta$ in a narrow region around
$\beta=10.635$ we find very clear tunnelling transitions 
on all lattice sizes. These become more rare as $L\uparrow$,
just as one would expect for a first order transition. In
Fig.\ref{fig_history4} we illustrate this with the time histories of 
$|{\bar l}_p|$ on the $L=14$ and $L=20$ lattices at 
$\beta=10.635$. 

\begin	{figure}[p]
\begin	{center}
\leavevmode
\epsfig{figure=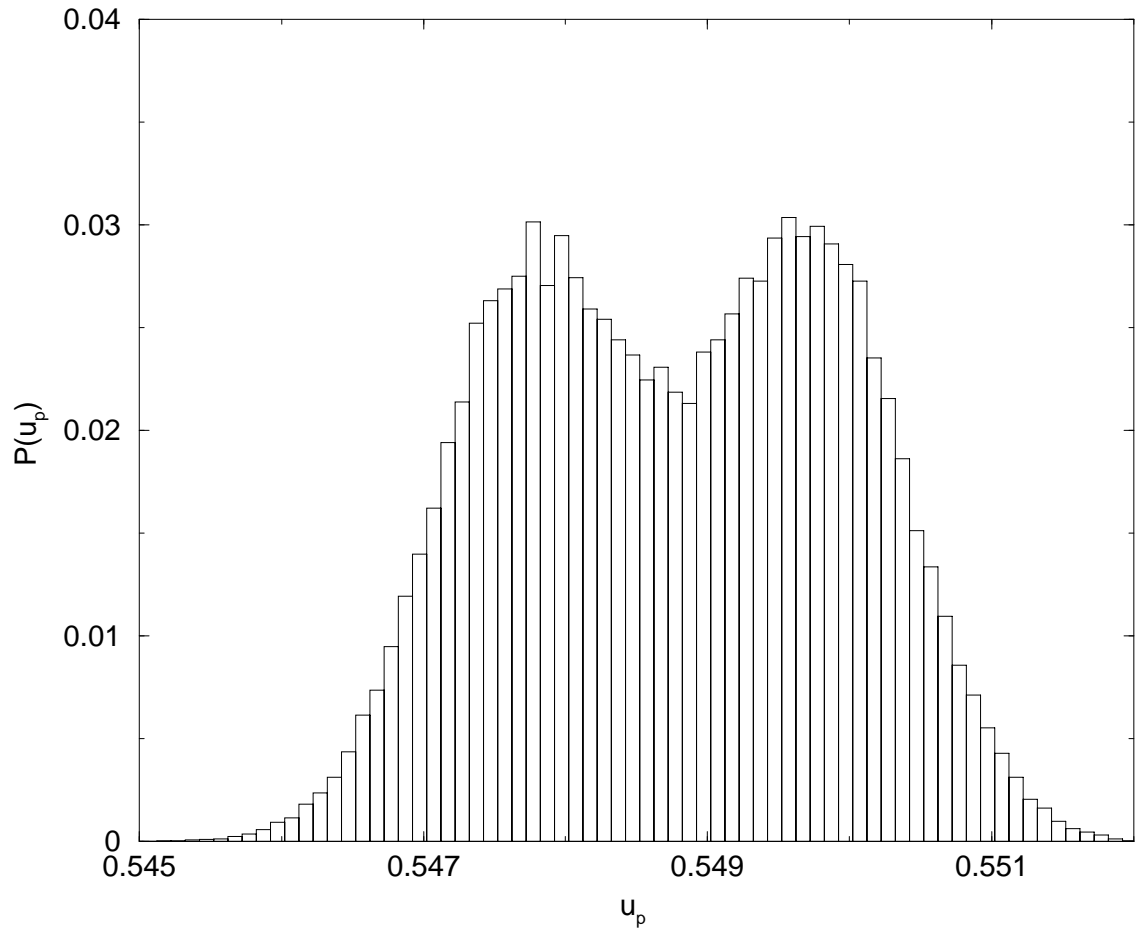, angle=270, width=15cm} 
\end	{center}
\vskip 0.15in
\caption{A histogram of the values of the average plaquette, for
the $20^3 5$ lattice at $\beta_c$ in SU(4).}
\label{fig_energy}
\end 	{figure}

\begin	{figure}[p]
\begin	{center}
\leavevmode
\epsfig{figure=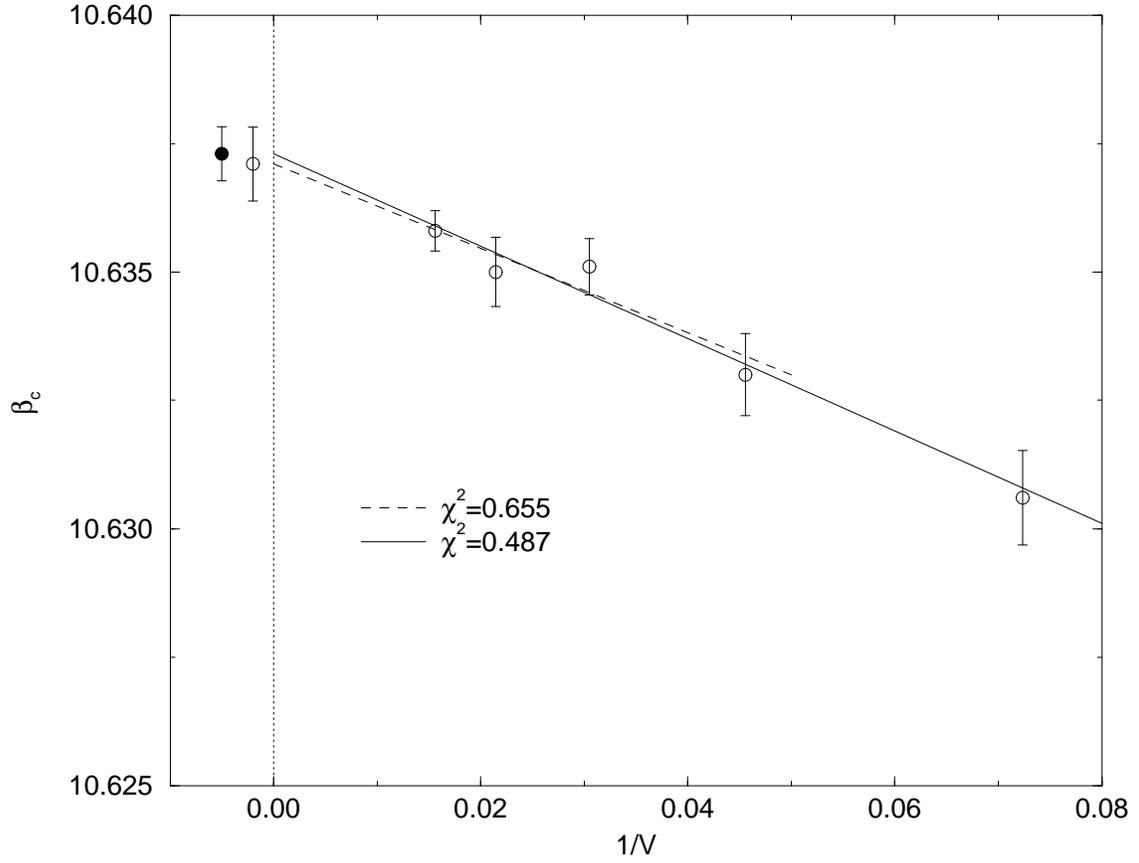, angle=270, width=15cm} 
\end	{center}
\vskip 0.15in
\caption{The critical value of $\beta$ plotted against the
inverse spatial volume, $V$, expressed in units of the temperature, $T$.
On $L^3 5$ lattices in SU(4). The straight lines are the 
extrapolations to infinite volume, one excluding the $L=12$
value. The corresponding extrapolated values are shown slightly to the
left of $1/V=0$.}
\label{fig_bcrit4}
\end 	{figure}

We define a normalised Polyakov loop susceptibility    
\begin{equation}
{{\chi_l}\over{V}} 
= 
\langle |{\bar l}_p|^2 \rangle - {\langle |{\bar l}_p| \rangle}^2
\label{C1}
\end{equation}
where we expect
\begin{equation}
\lim_{T\to T_c}{\chi_l} 
\stackrel{V\to \infty}{\propto}
\cases{V, &if 1'st order \cr
V^{\gamma}, &if 2'nd order \cr} 
\label{C2}
\end{equation}
with $\gamma$ a constant that depends on both the space-time
dimension and the critical exponents of the transition.
For each value of $L$ we calculate $\chi_l$ as a function of
$\beta$, determine the value at which it has a maximum
and use that as our estimate of $\beta_c$, and hence $T_c$. 
This is a good,
although not unique, criterion irrespective of whether
the transition is first or second order. To identify the
maximum we interpolate between the values of $\beta$ at 
which we have performed calculations (usually 
there are three that have
sufficient tunnellings to be useful) using a standard
reweighting technique
\cite{reweight}.
Fig.\ref{fig_energy} shows a histogram of the plaquette, averaged
over the lattice volume, for the $20^3 5$ lattices
reweighted to the critical beta-value. It
clearly shows the double peaking that is characteristic of
a first order transition.
In Fig.\ref{fig_bcrit4} we plot our values of $\beta_c$ against the 
inverse volume expressed in units of $T$. We note that if the transition
is first order, as our's clearly is, then there exist arguments 
\cite{Tsu3}
that with this way of determining $\beta_c$ the leading finite-$V$ 
correction should be 
\begin{equation}
\beta_c(V)
= 
\beta_c(\infty) - {h\over{VT^3}}.
\label{C3}
\end{equation}
Here $h$ is independent of the lattice action and of the lattice
spacing $a$ 
up to lattice corrections that will be small if $a$ is small.
We observe from Fig.\ref{fig_bcrit4} that all our values of 
$\beta_c$ are consistent
with eqn(\ref{C3}). Moreover the value of $h$ we thus extract
\begin{equation}
h = 0.09\pm 0.02 
\label{C4}
\end{equation}
is close to the SU(3) value, $h \simeq 0.1$
\cite{Tsu3}.
We take this to imply that $h$ depends weakly on $N$
and will use the value in eqn(\ref{C4}) to extrapolate to
$V=\infty$ in the SU(6) case where we do not perform
an explicit finite volume study. We remark that if $h$ is
indeed independent of $N$, then this implies that the volume 
dependence of $T_c(V)/T_c(\infty)$ is $O(1/N^2)$, {\it i.e.}
there is no volume dependence in the $N\to\infty$ limit.

\begin	{figure}[p]
\begin	{center}
\leavevmode
\epsfig{figure=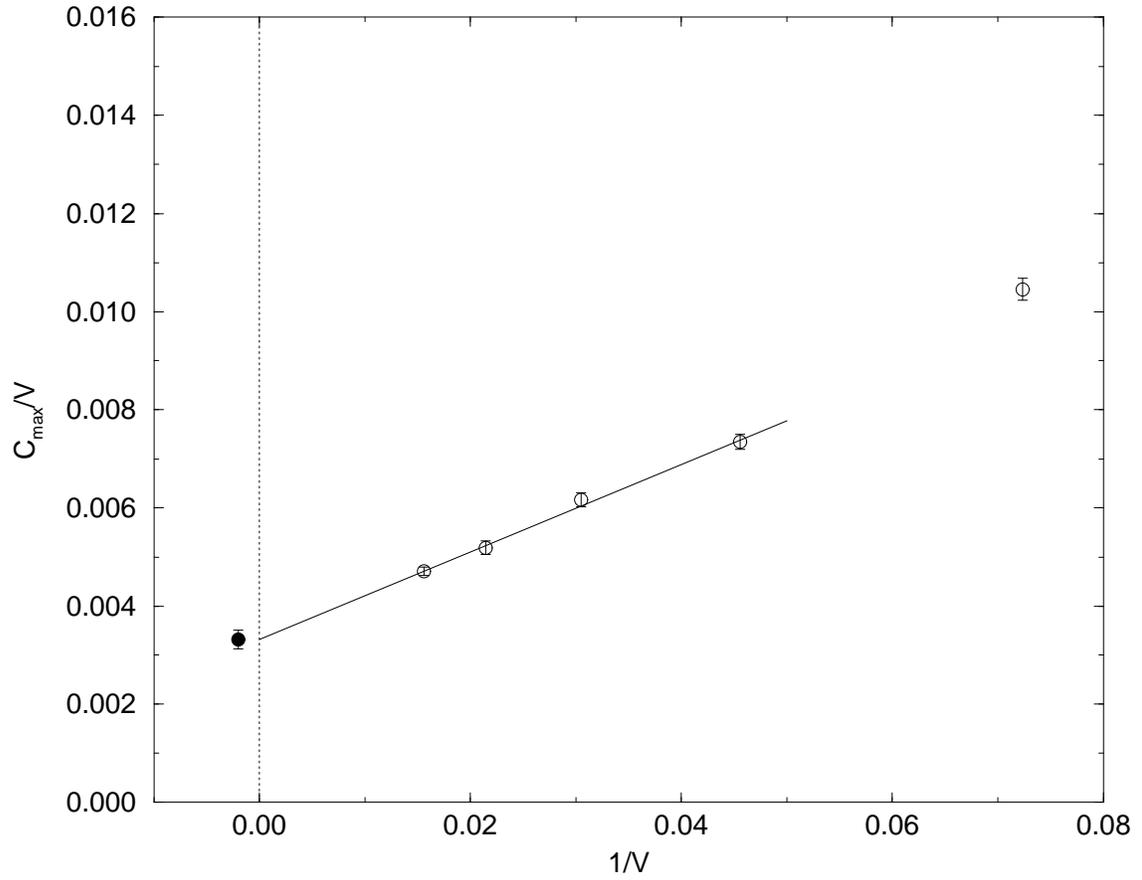, angle=270, width=15cm} 
\end	{center}
\vskip 0.15in
\caption{The specific heat at the critical value of $\beta$, 
$C(\beta_c,V)$, normalised to the spatial volume $V$ and plotted
against $1/V$, with $V$ in units of $T$. The intercept 
($\bullet$) at $V=\infty$ provides a measure of the latent heat. 
For SU(4) and $L_t=5$.}
\label{fig_susc4}
\end 	{figure}

A quantity analogous to $\chi_l$, but with plaquettes
replacing the Polyakov loop, is the specific heat $C(\beta)$:
\begin{equation}
{1\over{\beta^2}} C(\beta) 
= 
{\partial\over{\partial\beta}} 
\langle \bar{u}_p  \rangle
=
{N_p}
\langle \bar{u}_p^2 \rangle 
- 
{N_p}
{\langle \bar{u}_p  \rangle }^2
\label{C5}
\end{equation}
where $\bar{u}_p$ is the average value for a given lattice field
of the plaquette $u_p\equiv{\mathrm {ReTr}} U_p/N$, and $N_p=6L^3L_t$
is the total number of plaquettes. 
In Fig.\ref{fig_susc4} we plot $C(\beta_c)/V$ against $1/V$. We observe 
that it is certainly consistent with going to a finite value as 
$V \to \infty$, as one would expect for a first order transition.
We show in Fig.\ref{fig_susc4} a fit with a leading linear 
correction in $1/V$ of the kind that one would expect irrespective
of whether the transition was first or second order. For a
first order transition, as we appear to have here, 
we can obtain the latent heat, $L_h$, from
the intercept
\begin{equation}
\lim_{V\to\infty} {1\over{\beta_c^2 N_p}} C(\beta_c)
=
{1\over{4}}
(\langle \bar{u}_{p,c} \rangle - \langle\bar{u}_{p,d}\rangle)^2
=
{1\over{4}} L^2_h
\label{C6}
\end{equation}
where $\bar{u}_{p,c}$ and $\bar{u}_{p,d}$ are the average plaquette
values at $\beta=\beta_c$  in the confined and deconfined phases 
respectively. From the fit in Fig.\ref{fig_susc4} we obtain
\begin{equation}
L_h = 0.00197(5).
\label{C7}
\end{equation}
We note that if we take the  $L_t=4$ SU(3) latent heat in
\cite{ABS}
and naively scale it to $L_t=5$, we obtain
$L_h(N=3) \simeq 0.0013$ which is substantially smaller
than our above SU(4) value. This shows explicitly that the SU(4)
transition is more strongly first-order than the SU(3) one.

\begin	{figure}[p]
\begin	{center}
\leavevmode
\setlength{\unitlength}{0.240900pt}
\ifx\plotpoint\undefined\newsavebox{\plotpoint}\fi
\sbox{\plotpoint}{\rule[-0.200pt]{0.400pt}{0.400pt}}%
\begin{picture}(1500,1800)(0,0)
\font\gnuplot=cmr10 at 12pt
\gnuplot
\sbox{\plotpoint}{\rule[-0.200pt]{0.400pt}{0.400pt}}%
\put(425.0,250.0){\rule[-0.200pt]{4.818pt}{0.400pt}}
\put(400,250){\makebox(0,0)[r]{\ \ {$0$}}}
\put(1405.0,250.0){\rule[-0.200pt]{4.818pt}{0.400pt}}
\put(425.0,550.0){\rule[-0.200pt]{4.818pt}{0.400pt}}
\put(400,550){\makebox(0,0)[r]{\ \ {$0.0005$}}}
\put(1405.0,550.0){\rule[-0.200pt]{4.818pt}{0.400pt}}
\put(425.0,850.0){\rule[-0.200pt]{4.818pt}{0.400pt}}
\put(400,850){\makebox(0,0)[r]{\ \ {$0.001$}}}
\put(1405.0,850.0){\rule[-0.200pt]{4.818pt}{0.400pt}}
\put(425.0,1150.0){\rule[-0.200pt]{4.818pt}{0.400pt}}
\put(400,1150){\makebox(0,0)[r]{\ \ {$0.0015$}}}
\put(1405.0,1150.0){\rule[-0.200pt]{4.818pt}{0.400pt}}
\put(425.0,1450.0){\rule[-0.200pt]{4.818pt}{0.400pt}}
\put(400,1450){\makebox(0,0)[r]{\ \ {$0.002$}}}
\put(1405.0,1450.0){\rule[-0.200pt]{4.818pt}{0.400pt}}
\put(425.0,250.0){\rule[-0.200pt]{0.400pt}{4.818pt}}
\put(425,200){\makebox(0,0){\ {$0$}}}
\put(425.0,1730.0){\rule[-0.200pt]{0.400pt}{4.818pt}}
\put(568.0,250.0){\rule[-0.200pt]{0.400pt}{4.818pt}}
\put(568,200){\makebox(0,0){\ {$5$}}}
\put(568.0,1730.0){\rule[-0.200pt]{0.400pt}{4.818pt}}
\put(711.0,250.0){\rule[-0.200pt]{0.400pt}{4.818pt}}
\put(711,200){\makebox(0,0){\ {$10$}}}
\put(711.0,1730.0){\rule[-0.200pt]{0.400pt}{4.818pt}}
\put(854.0,250.0){\rule[-0.200pt]{0.400pt}{4.818pt}}
\put(854,200){\makebox(0,0){\ {$15$}}}
\put(854.0,1730.0){\rule[-0.200pt]{0.400pt}{4.818pt}}
\put(996.0,250.0){\rule[-0.200pt]{0.400pt}{4.818pt}}
\put(996,200){\makebox(0,0){\ {$20$}}}
\put(996.0,1730.0){\rule[-0.200pt]{0.400pt}{4.818pt}}
\put(1139.0,250.0){\rule[-0.200pt]{0.400pt}{4.818pt}}
\put(1139,200){\makebox(0,0){\ {$25$}}}
\put(1139.0,1730.0){\rule[-0.200pt]{0.400pt}{4.818pt}}
\put(1282.0,250.0){\rule[-0.200pt]{0.400pt}{4.818pt}}
\put(1282,200){\makebox(0,0){\ {$30$}}}
\put(1282.0,1730.0){\rule[-0.200pt]{0.400pt}{4.818pt}}
\put(425.0,250.0){\rule[-0.200pt]{240.900pt}{0.400pt}}
\put(1425.0,250.0){\rule[-0.200pt]{0.400pt}{361.350pt}}
\put(425.0,1750.0){\rule[-0.200pt]{240.900pt}{0.400pt}}
\put(150,1300){\makebox(0,0){\Large{${\Delta s }$}}}
\put(900,25){\makebox(0,0){\large{$L$}}}
\put(425.0,250.0){\rule[-0.200pt]{0.400pt}{361.350pt}}
\put(768.0,1210.0){\rule[-0.200pt]{0.400pt}{86.724pt}}
\put(758.0,1210.0){\rule[-0.200pt]{4.818pt}{0.400pt}}
\put(758.0,1570.0){\rule[-0.200pt]{4.818pt}{0.400pt}}
\put(825.0,1210.0){\rule[-0.200pt]{0.400pt}{57.816pt}}
\put(815.0,1210.0){\rule[-0.200pt]{4.818pt}{0.400pt}}
\put(815.0,1450.0){\rule[-0.200pt]{4.818pt}{0.400pt}}
\put(882.0,1270.0){\rule[-0.200pt]{0.400pt}{28.908pt}}
\put(872.0,1270.0){\rule[-0.200pt]{4.818pt}{0.400pt}}
\put(872.0,1390.0){\rule[-0.200pt]{4.818pt}{0.400pt}}
\put(994.0,1324.0){\rule[-0.200pt]{0.400pt}{14.454pt}}
\put(984.0,1324.0){\rule[-0.200pt]{4.818pt}{0.400pt}}
\put(984.0,1384.0){\rule[-0.200pt]{4.818pt}{0.400pt}}
\put(1339.0,1312.0){\rule[-0.200pt]{0.400pt}{14.454pt}}
\put(1329.0,1312.0){\rule[-0.200pt]{4.818pt}{0.400pt}}
\put(768,1390){\circle{18}}
\put(825,1330){\circle{18}}
\put(882,1330){\circle{18}}
\put(994,1354){\circle{18}}
\put(1339,1342){\circle{18}}
\put(1329.0,1372.0){\rule[-0.200pt]{4.818pt}{0.400pt}}
\put(939.0,1324.0){\rule[-0.200pt]{0.400pt}{23.126pt}}
\put(929.0,1324.0){\rule[-0.200pt]{4.818pt}{0.400pt}}
\put(929.0,1420.0){\rule[-0.200pt]{4.818pt}{0.400pt}}
\put(999.0,1348.0){\rule[-0.200pt]{0.400pt}{11.563pt}}
\put(989.0,1348.0){\rule[-0.200pt]{4.818pt}{0.400pt}}
\put(939,1372){\circle*{18}}
\put(999,1372){\circle*{18}}
\put(989.0,1396.0){\rule[-0.200pt]{4.818pt}{0.400pt}}
\end{picture}

\end	{center}
\vskip 0.15in
\caption{The difference of the average plaquette in the confining and
deconfining phases on $L^3 5$ SU(4) lattices plotted versus $L$.
For $\beta=10.635$ ($\circ$) and $\beta=10.637$ ($\bullet$).}
\label{fig_diffS4}
\end 	{figure}
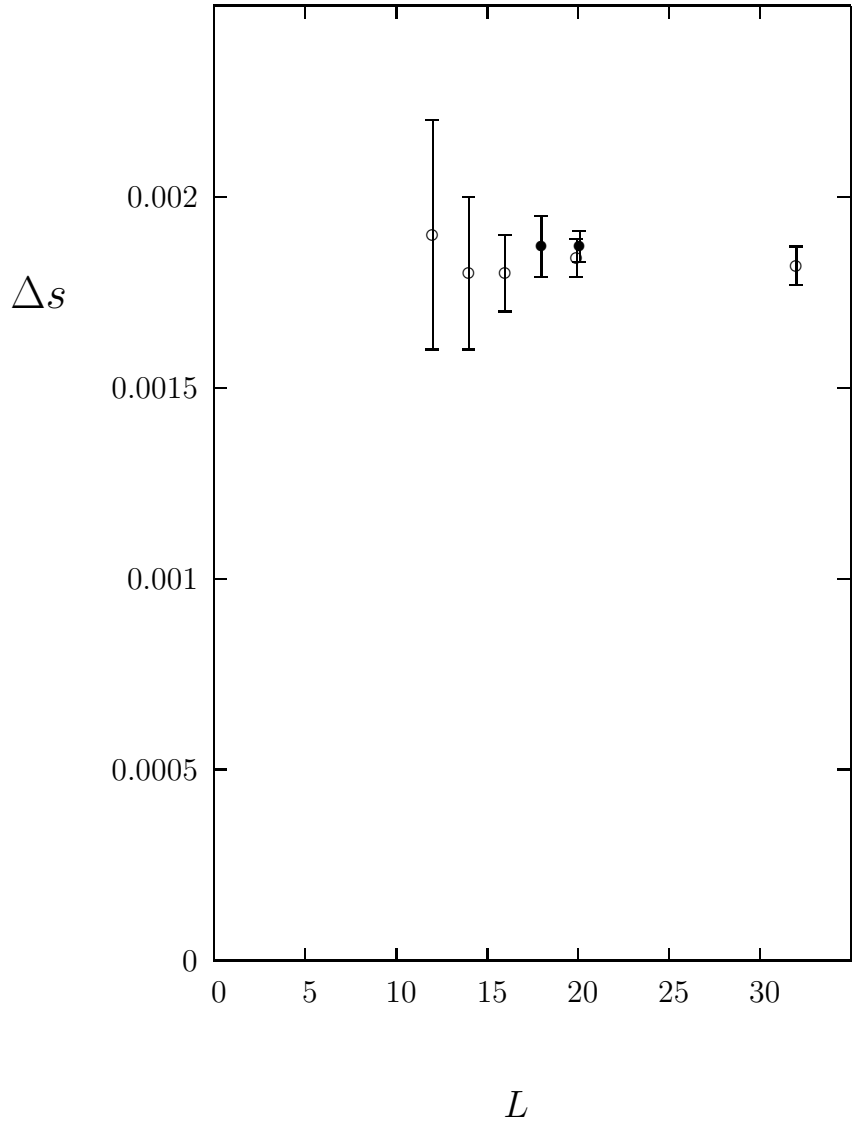

We have seen in Fig.\ref{fig_history4} that the value of ${\bar l}_p$ 
allows us to label the phase of a field configuration almost 
unambiguously. So for values of $\beta$ close to $\beta_c$, where
all phases are sampled, we can calculate the average plaquette in each 
phase separately and hence the difference $\Delta s$ between the
confined and deconfined phases. Calculated at $\beta_c$ this is the
finite-$V$ latent heat of the transition. There is some uncertainty in 
separating configurations in a single phase from those that are
tunnelling. This is a systematic error that we can estimate by
varying the cuts on ${\bar l}_p$, and we include it in our
final error estimates. In Fig.\ref{fig_diffS4} we show how $\Delta s$
varies with the spatial size, $L$, for those of our calculations that
are closest to $\beta_c$ (see Fig.\ref{fig_bcrit4}). We see that
the latent heat is at most weakly dependent on the spatial volume
and appears to have a finite $V=\infty$ limit, confirming
once again that this is a first order phase transition. The value
is pleasingly close to that in eqn(\ref{C7}), which has been obtained
under very different assumptions.

The value of $\Delta s$ provides a measure of how strong is
the transition. To express it in physical units we note
that if one separates the plaquette into an ultraviolet
perturbative piece and the gluon condensate
\cite{Gcondensate},
then the lattice renormalisation factor multiplying the
latter is $Z \simeq 1$. Thus we can express 
$\Delta s \equiv a^4 m^4_s$ where $m_s$ is a physical scale.
Noting that for $\beta\simeq\beta_c$ we have
$\Delta s \simeq 0.002$ and that $aT_c = 1/L_t = 0.2$
we see that $m_s \simeq T_c$; that is to say, a typical
dynamical scale indicating that the strength of this
first order transition is quite ordinary. We remark
that our preliminary calculations with $L_t = 6$ indicate that
lattice corrections are small and that this statement
holds in the continuum limit.

From the fit in Fig.\ref{fig_bcrit4} we obtain the $V=\infty$
critical value $\beta_c = 10.63709(72)$. Interpolating previously
calculated values of the string tension 
\cite{blmt-glue}
to $\beta_c$, and using $T_c=1/5a(\beta_c)$, we obtain
\begin{equation}
{{T_c}\over{\sqrt{\sigma}}}
=
0.6024 \pm 0.0045  \ \ \ \ \ \ \ {\mathrm {at}} \ a=1/5T_c
\label{C8}
\end{equation}
where the bulk of the error comes from the string tension.
If lattice corrections are very small, as is found to be the
case in SU(3)
\cite{Tsu3},
then eqn(\ref{C8}) should provide a good estimate of the value
in the continuum limit. In fact our very preliminary $L_t=6$ 
calculations give a $V=\infty$ value $\beta_c=10.780(10)$ which 
translates to $T_c/\sqrt{\sigma} = 0.597(8)$, where the major part
of the error comes from the uncertainty in $\beta_c$. This, taken 
together with the value in eqn(\ref{C7}), gives us the extrapolated 
continuum value
\begin{equation}
\lim_{a\to 0}{{T_c}\over{\sqrt{\sigma}}}
=
0.584 \pm 0.030 .
\label{C9}
\end{equation}
In making such
an extrapolation we need to assume the dominance of the leading
$O(a^2\sigma)$ lattice correction, and this is justified by the
observed dominance of this correction for other physical
quantities
\cite{blmt-glue,blmt-kstring}.
\section{$T_c$ in SU(6)}

Our high statistics SU(6) calculations have been performed 
on $16^3 6$ lattices at values of $\beta$ close to
$\beta=24.845$. In physical units this is like a $14^3$
lattice with $L_t=5$ and although this is not large, we
have seen in our SU(4) calculations (see Figs.1-4) that the 
physics of such volumes is quite close to the thermodynamic limit.

\begin	{figure}[p]
\begin	{center}
\leavevmode
\epsfig{figure=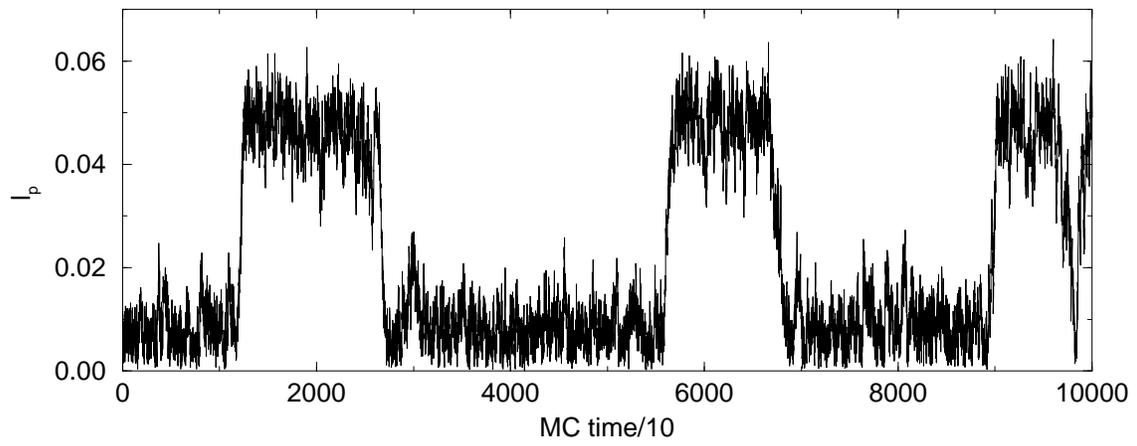, angle=270, width=15cm} 
\end	{center}
\vskip 0.15in
\caption{The modulus of the average value of the Polyakov loop 
for a sequence of $16^3 6$ SU(6) field 
configurations at $\beta=24.845$.}
\label{fig_history6}
\end 	{figure}

\begin	{figure}[p]
\begin	{center}
\leavevmode
\epsfig{figure=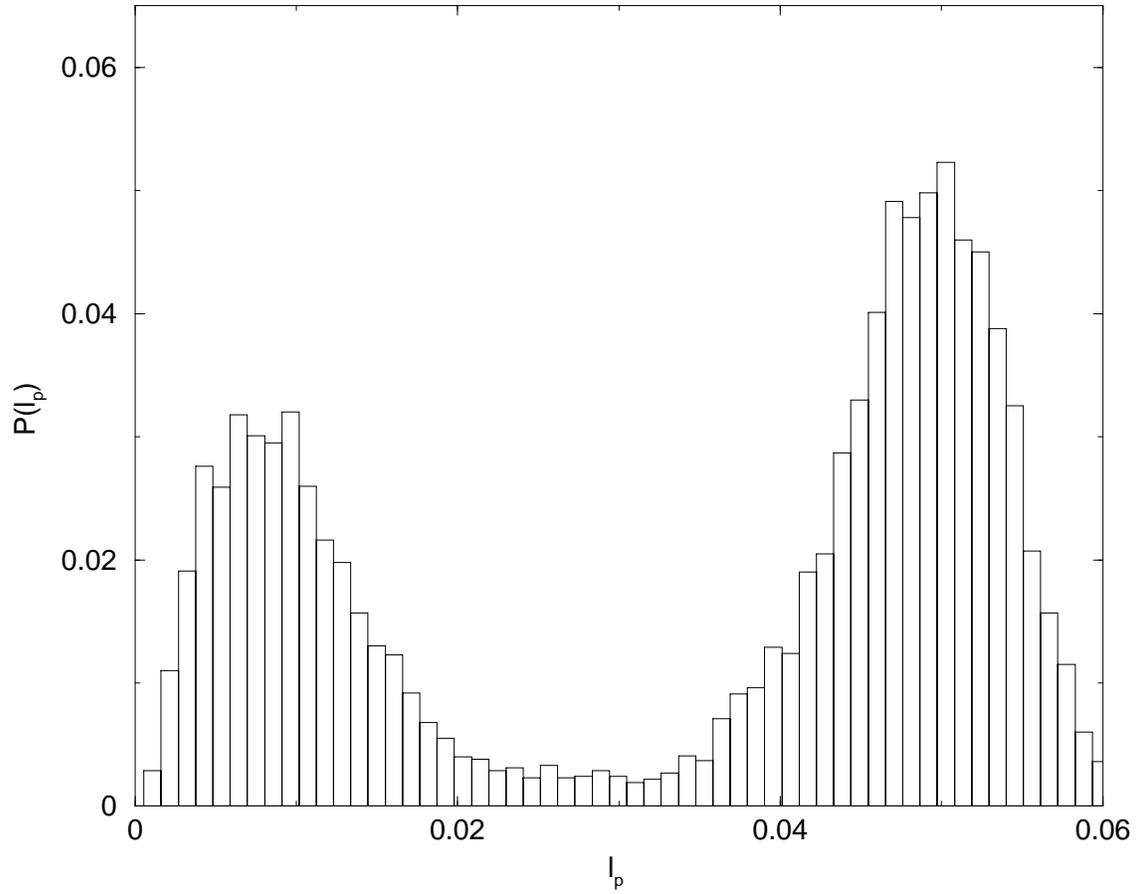, angle=270, width=15cm} 
\end	{center}
\vskip 0.15in
\caption{A histogram of the values the modulus of the Polyakov loop,
${\bar l}_p$, for the SU(6) calculation at  $\beta=24.85$.}
\label{fig_histo6}
\end 	{figure}

In Fig.\ref{fig_history6} we plot the values of $|{\bar l}_p|$
for the sequence of configurations that we obtain at
$\beta=24.845$. We see well defined 
tunnelling between confined and deconfined phases, 
characteristic of a first order transition. This is
further illustrated in Fig.\ref{fig_histo6} which displays the
very clear separation between the phases --
this time at $\beta=24.85$. 
Using the reweighting technique we extract a critical value 
$\beta_c=24.850(3)$. We extrapolate this to $V=\infty$ using
eqn(\ref{C3}) with the value of $h$ in eqn(\ref{C4}). Recall
that this is motivated by the fact that one finds $h$ to
have the same value in SU(3) and SU(4). Doing so we
obtain $\lim_{V\to\infty} \beta_c = 24.855\pm0.003$.
This can be translated into a value of $T_c$ just as in
the case of SU(4), using the values of the SU(6) string tension
calculated in
\cite{Pisa1},
to give
\begin{equation}
{{T_c}\over{\sqrt{\sigma}}}
=
0.588 \pm 0.002  \ \ \ \ \ \ \ {\mathrm {at}} \ a=1/6T_c. 
\label{D1}
\end{equation}
Just as in SU(4), we would expect this to provide a good estimate of
the continuum value. We have in addition performed some preliminary 
calculations on $16^3 5$ lattices, increasing and decreasing
$\beta$ while crossing the deconfining phase transition. From the
average plaquette values it is clear that these calculations
are all on the weak coupling side of the nearby bulk transition 
and can therefore be used for a continuum extrapolation. (Whether 
this remains so in a much higher statistics calculation and for
other volumes is something we are in the process of investigating
\cite{blmtuw}.)
These preliminary $L_t=5$ calculations provide, after
extrapolation to $V=\infty$, an estimate $\beta_c = 24.519(33)$ which
translates to $T_c/\sqrt{\sigma} = 0.580(12)$. Taken together with
with the value in eqn(\ref{D1}) this gives us the continuum value
\begin{equation}
\lim_{a\to 0}{{T_c}\over{\sqrt{\sigma}}}
=
0.605 \pm 0.026 .
\label{D2}
\end{equation}
As in the case of SU(4) we need to assume the dominance of the leading
$O(a^2\sigma)$ lattice correction, and the justification
is the same as it was there
\cite{Pisa1}.

As for the strength of the first order transition, we remark
that our value of $\Delta s$ is larger than our (preliminary)
value on a directly comparable $16^3 6$ SU(4)
calculation, showing that the transition is certainly not 
weakening as $N\to\infty$.

\section{Discussion}

Our careful finite size study in SU(4) for $aT = 1/5$ reveals 
quite clearly that the deconfining phase transition is first 
order with a latent heat that is not particularly small.
This confirms previous work
\cite{winoh,gavai}.
Our SU(6) calculations do not involve a finite size study,
but have been performed on a spatial volume that, on the
basis of our SU(4) calculations, should be large enough
for our purposes. The calculations were performed closer
to the continuum limit, $aT = 1/6$, in order to avoid 
any possible confusion with a nearby strongly first order
bulk transition. On our $16^3 6$ lattices we found
very clear, if rare, tunnelling between the confining
and deconfining phases, characteristic of a typical
first order transition. The extrapolation to the thermodynamic 
limit was made using the dependence on $V$ found in SU(4). This 
can be justified by the fact that the SU(4) volume dependence is the 
same as found in SU(3), suggesting that it depends at most weakly 
on $N$. The SU(6) transition appears to be at least as strong
as the SU(4) one, which tells us that the SU($N=\infty$)
transition is first order and not particularly weak.

\begin	{figure}[p]
\begin	{center}
\leavevmode
\setlength{\unitlength}{0.240900pt}
\ifx\plotpoint\undefined\newsavebox{\plotpoint}\fi
\sbox{\plotpoint}{\rule[-0.200pt]{0.400pt}{0.400pt}}%
\begin{picture}(1500,1800)(0,0)
\font\gnuplot=cmr10 at 12pt
\gnuplot
\sbox{\plotpoint}{\rule[-0.200pt]{0.400pt}{0.400pt}}%
\put(350.0,250.0){\rule[-0.200pt]{4.818pt}{0.400pt}}
\put(325,250){\makebox(0,0)[r]{\ \ {$0$}}}
\put(1405.0,250.0){\rule[-0.200pt]{4.818pt}{0.400pt}}
\put(350.0,481.0){\rule[-0.200pt]{4.818pt}{0.400pt}}
\put(325,481){\makebox(0,0)[r]{\ \ {$0.2$}}}
\put(1405.0,481.0){\rule[-0.200pt]{4.818pt}{0.400pt}}
\put(350.0,712.0){\rule[-0.200pt]{4.818pt}{0.400pt}}
\put(325,712){\makebox(0,0)[r]{\ \ {$0.4$}}}
\put(1405.0,712.0){\rule[-0.200pt]{4.818pt}{0.400pt}}
\put(350.0,942.0){\rule[-0.200pt]{4.818pt}{0.400pt}}
\put(325,942){\makebox(0,0)[r]{\ \ {$0.6$}}}
\put(1405.0,942.0){\rule[-0.200pt]{4.818pt}{0.400pt}}
\put(350.0,1173.0){\rule[-0.200pt]{4.818pt}{0.400pt}}
\put(325,1173){\makebox(0,0)[r]{\ \ {$0.8$}}}
\put(1405.0,1173.0){\rule[-0.200pt]{4.818pt}{0.400pt}}
\put(350.0,1404.0){\rule[-0.200pt]{4.818pt}{0.400pt}}
\put(325,1404){\makebox(0,0)[r]{\ \ {$1$}}}
\put(1405.0,1404.0){\rule[-0.200pt]{4.818pt}{0.400pt}}
\put(350.0,1635.0){\rule[-0.200pt]{4.818pt}{0.400pt}}
\put(325,1635){\makebox(0,0)[r]{\ \ {$1.2$}}}
\put(1405.0,1635.0){\rule[-0.200pt]{4.818pt}{0.400pt}}
\put(350.0,250.0){\rule[-0.200pt]{0.400pt}{4.818pt}}
\put(350,200){\makebox(0,0){\ {$0$}}}
\put(350.0,1730.0){\rule[-0.200pt]{0.400pt}{4.818pt}}
\put(515.0,250.0){\rule[-0.200pt]{0.400pt}{4.818pt}}
\put(515,200){\makebox(0,0){\ {$0.05$}}}
\put(515.0,1730.0){\rule[-0.200pt]{0.400pt}{4.818pt}}
\put(681.0,250.0){\rule[-0.200pt]{0.400pt}{4.818pt}}
\put(681,200){\makebox(0,0){\ {$0.1$}}}
\put(681.0,1730.0){\rule[-0.200pt]{0.400pt}{4.818pt}}
\put(846.0,250.0){\rule[-0.200pt]{0.400pt}{4.818pt}}
\put(846,200){\makebox(0,0){\ {$0.15$}}}
\put(846.0,1730.0){\rule[-0.200pt]{0.400pt}{4.818pt}}
\put(1012.0,250.0){\rule[-0.200pt]{0.400pt}{4.818pt}}
\put(1012,200){\makebox(0,0){\ {$0.2$}}}
\put(1012.0,1730.0){\rule[-0.200pt]{0.400pt}{4.818pt}}
\put(1177.0,250.0){\rule[-0.200pt]{0.400pt}{4.818pt}}
\put(1177,200){\makebox(0,0){\ {$0.25$}}}
\put(1177.0,1730.0){\rule[-0.200pt]{0.400pt}{4.818pt}}
\put(1342.0,250.0){\rule[-0.200pt]{0.400pt}{4.818pt}}
\put(1342,200){\makebox(0,0){\ {$0.3$}}}
\put(1342.0,1730.0){\rule[-0.200pt]{0.400pt}{4.818pt}}
\put(350.0,250.0){\rule[-0.200pt]{258.967pt}{0.400pt}}
\put(1425.0,250.0){\rule[-0.200pt]{0.400pt}{361.350pt}}
\put(350.0,1750.0){\rule[-0.200pt]{258.967pt}{0.400pt}}
\put(150,1300){\makebox(0,0){\Large{${{T_c} \over{\surd\sigma} }$}}}
\put(862,25){\makebox(0,0){\large{$1/N^2$}}}
\put(350.0,250.0){\rule[-0.200pt]{0.400pt}{361.350pt}}
\put(1177.0,1023.0){\rule[-0.200pt]{0.400pt}{11.081pt}}
\put(1167.0,1023.0){\rule[-0.200pt]{4.818pt}{0.400pt}}
\put(1167.0,1069.0){\rule[-0.200pt]{4.818pt}{0.400pt}}
\put(717.0,971.0){\rule[-0.200pt]{0.400pt}{2.891pt}}
\put(707.0,971.0){\rule[-0.200pt]{4.818pt}{0.400pt}}
\put(707.0,983.0){\rule[-0.200pt]{4.818pt}{0.400pt}}
\put(557.0,889.0){\rule[-0.200pt]{0.400pt}{16.622pt}}
\put(547.0,889.0){\rule[-0.200pt]{4.818pt}{0.400pt}}
\put(547.0,958.0){\rule[-0.200pt]{4.818pt}{0.400pt}}
\put(442.0,918.0){\rule[-0.200pt]{0.400pt}{14.454pt}}
\put(432.0,918.0){\rule[-0.200pt]{4.818pt}{0.400pt}}
\put(1177,1046){\circle*{12}}
\put(717,977){\circle*{12}}
\put(557,924){\circle*{12}}
\put(442,948){\circle*{12}}
\put(432.0,978.0){\rule[-0.200pt]{4.818pt}{0.400pt}}
\sbox{\plotpoint}{\rule[-0.500pt]{1.000pt}{1.000pt}}%
\put(350,921){\usebox{\plotpoint}}
\put(350.00,921.00){\usebox{\plotpoint}}
\put(370.54,923.87){\usebox{\plotpoint}}
\put(390.95,927.59){\usebox{\plotpoint}}
\put(411.49,930.36){\usebox{\plotpoint}}
\put(431.99,933.54){\usebox{\plotpoint}}
\put(452.47,936.81){\usebox{\plotpoint}}
\put(473.00,939.73){\usebox{\plotpoint}}
\put(493.55,942.46){\usebox{\plotpoint}}
\put(513.99,946.09){\usebox{\plotpoint}}
\put(534.53,948.91){\usebox{\plotpoint}}
\put(555.04,951.91){\usebox{\plotpoint}}
\put(575.47,955.54){\usebox{\plotpoint}}
\put(596.02,958.28){\usebox{\plotpoint}}
\put(616.51,961.55){\usebox{\plotpoint}}
\put(637.04,964.46){\usebox{\plotpoint}}
\put(657.53,967.64){\usebox{\plotpoint}}
\put(678.08,970.38){\usebox{\plotpoint}}
\put(698.49,974.14){\usebox{\plotpoint}}
\put(719.02,977.00){\usebox{\plotpoint}}
\put(739.56,979.87){\usebox{\plotpoint}}
\put(760.10,982.74){\usebox{\plotpoint}}
\put(780.52,986.37){\usebox{\plotpoint}}
\put(801.07,989.10){\usebox{\plotpoint}}
\put(821.55,992.41){\usebox{\plotpoint}}
\put(842.04,995.61){\usebox{\plotpoint}}
\put(862.57,998.47){\usebox{\plotpoint}}
\put(883.12,1001.20){\usebox{\plotpoint}}
\put(903.54,1004.92){\usebox{\plotpoint}}
\put(924.09,1007.65){\usebox{\plotpoint}}
\put(944.59,1010.78){\usebox{\plotpoint}}
\put(965.04,1014.28){\usebox{\plotpoint}}
\put(985.59,1017.02){\usebox{\plotpoint}}
\put(1006.05,1020.41){\usebox{\plotpoint}}
\put(1026.59,1023.33){\usebox{\plotpoint}}
\put(1047.10,1026.38){\usebox{\plotpoint}}
\put(1067.65,1029.12){\usebox{\plotpoint}}
\put(1088.04,1033.00){\usebox{\plotpoint}}
\put(1108.59,1035.74){\usebox{\plotpoint}}
\put(1129.11,1038.74){\usebox{\plotpoint}}
\put(1149.64,1041.60){\usebox{\plotpoint}}
\put(1170.08,1045.11){\usebox{\plotpoint}}
\put(1190.64,1047.84){\usebox{\plotpoint}}
\put(1211.10,1051.28){\usebox{\plotpoint}}
\put(1231.61,1054.32){\usebox{\plotpoint}}
\put(1252.13,1057.21){\usebox{\plotpoint}}
\put(1272.68,1059.97){\usebox{\plotpoint}}
\put(1293.11,1063.66){\usebox{\plotpoint}}
\put(1313.63,1066.53){\usebox{\plotpoint}}
\put(1334.13,1069.65){\usebox{\plotpoint}}
\put(1354.60,1073.02){\usebox{\plotpoint}}
\put(1375.15,1075.76){\usebox{\plotpoint}}
\put(1395.58,1079.33){\usebox{\plotpoint}}
\put(1416.12,1082.19){\usebox{\plotpoint}}
\put(1425,1083){\usebox{\plotpoint}}
\end{picture}

\end	{center}
\vskip 0.15in
\caption{The deconfining temperature, $T_c$, in units of the
string tension, $\sigma$ plotted versus $1/N^2$. The
straight line is the large-$N$ extrapolation with a
leading $O(1/N^2)$ correction.}
\label{fig_TsuN}
\end 	{figure}
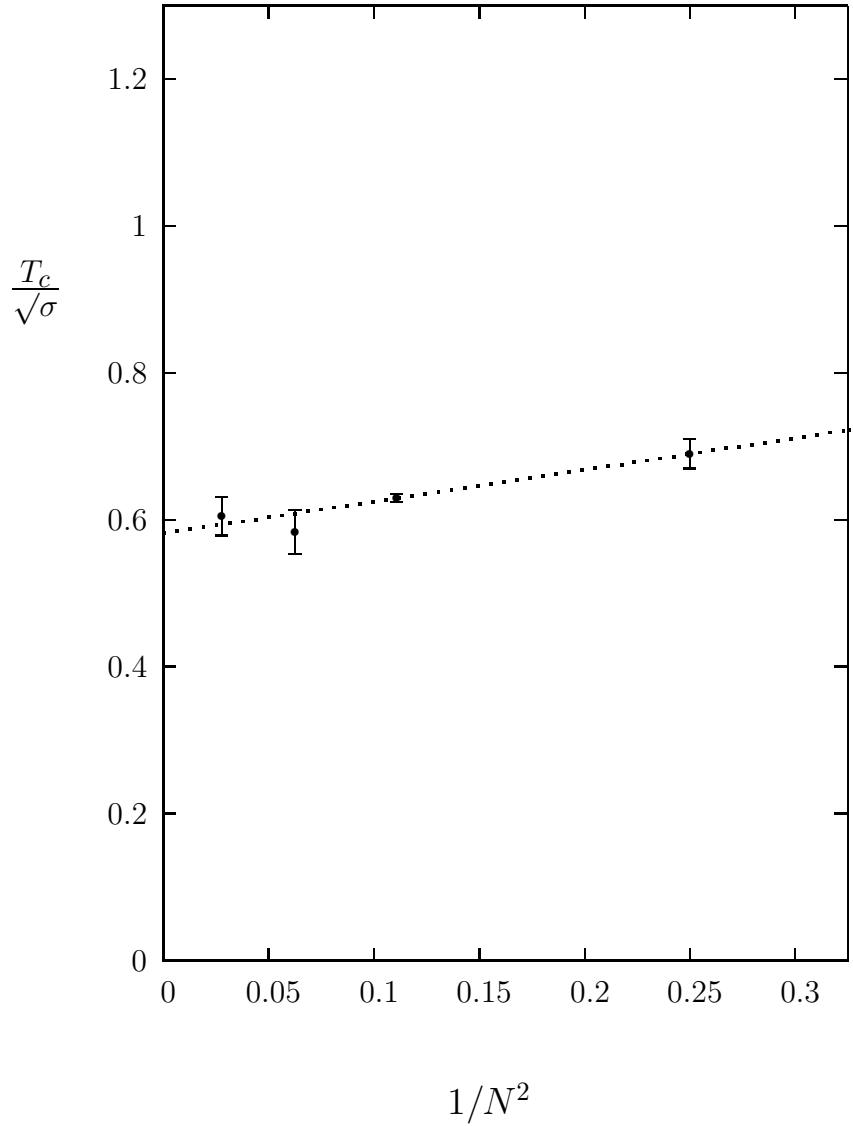

Our preliminary calculations at other values of $a$
suggest that the values of $T_c$ in eqns(\ref{C7},\ref{D1})
are indeed close to the continuum values. This is
not surprising given the very small lattice corrections
observed in SU(3) calculations with the same lattice action
\cite{Tsu3}.
If we take these SU(3) values with older SU(2) ones
\cite{Tsu2}
and plot them against $1/N^2$ together with our
SU(4) and SU(6) values, as in Fig.\ref{fig_TsuN},
we see that we can describe the values for all $N$
with just a leading large-$N$ correction
\begin{equation}
{{T_c}\over{\sqrt{\sigma}}}
=
0.582(15) + {{0.43(13)}\over{N^2}},
\label{E1}
\end{equation}
even though the SU(2) transition is second order, the
SU(3) is weakly first order and the SU(4) and SU(6)
transitions are `normal' first order transitions.
Our ongoing calculations will significantly reduce the
errors; both by providing accurate values of
$\beta_c$ at $L_t = 5,6$ for SU(6) and SU(4) respectively,
and because we intend to obtain more accurate values of
the string tension at the critical values of $\beta$
corresponding to $L_t=5$, where the error on our string 
tension extrapolation/interpolation is particularly large.
In addition our SU(3) calculations will use the same methods,
and so a comparison with larger $N$ should provide direct 
evidence for the strengthening of the transition as $N$
increases.

There are of course many other interesting properties of
the phase transition, some of which we have been
studying. We will report on this work elsewhere
\cite{blmtuw}.
For example, topological fluctuations are suppressed across 
$T_c$ and we find that this effect becomes dramatically stronger 
at larger $N$. Small instantons are explicitly suppressed
as $N\uparrow$ while larger instantons are suppressed
as $T\uparrow$ in the deconfined phase. It appears that $T_c$ is 
large enough for these effects to `cross' at $T_c$ 
at large $N$ so that there is essentially
no topology left for $T>T_c$. We also calculate the mass 
gap in the neighbourhood of the transition, separately for the
confined and deconfined phases, and explicitly confirm that
it does not vanish at $T_c$. We calculate the spatial string tensions
of $k$-strings near $T_c$ and confirm approximate Casimir scaling
\cite{blmt-kstring}
in both phases, although the string tensions are significantly
different in the two phases (at the same value of $\beta$).
Comparing $k=1$ and $k=2$ Polyakov loops we see that 
deconfinement appears to occur for both at the same value of $T_c$.
And our finite volume study in SU(6) should determine
whether our claim in this paper, that finite volume
effects disappear as $1/N^2$, is indeed correct.
Finally our extra calculations at other other values of $a$
will enable us to improve upon the calculations reported herein.

\section*{Acknowledgments}
These calculations were carried out on Alpha Compaq workstations 
in Oxford Theoretical Physics, funded by PPARC and EPSRC grants.
UW has been supported by a PPARC fellowship, and BL by a
EU Marie Sklodowska-Curie postdoctoral fellowship.

\newpage

\clearpage

\end{document}